\documentclass[man,fignum]{apa} 

\usepackage{amssymb}
\usepackage{subfigure,graphicx,amstext}
\usepackage{amsthm,url}
\usepackage{color}
\newtheorem{theorem}{Theorem}

\title{Algorithmic Complexity for Short Binary Strings Applied to Psychology: A Primer}
\fourauthors{Nicolas Gauvrit}{Hector Zenil}{Jean-Paul Delahaye}{Fernando Soler-Toscano}
\fouraffiliations{CHART (PARIS-reasoning), University of Paris VIII and EPHE, Paris, France}{Unit of Computational Medicine, Center for Molecular Medicine, Karolinska Institute, Stockholm, Sweden}{Laboratoire d'Informatique Fondamentale de Lille, France}{Grupo de L\'ogica, Lenguaje e Informaci\'on, Universidad de Sevilla, Spain.}

\note{Corresponding author: Nicolas Gauvrit, ngauvrit@me.com}

\abstract{
As human randomness production has come to be more closely  studied and used to assess executive functions (especially inhibition), many normative measures for assessing the degree to which a sequence is random-like have been suggested. However, each of these measures focus on one feature of randomness, leading researchers to have to use multiple measures. Although algorithmic complexity has been suggested as a means for overcoming this inconvenience, it has never been used because standard Kolmogorov complexity is inapplicable to short  strings (e.g. of length $l\leq50$), due to both computational and theoretical limitations. Here we describe a novel technique (the ``Coding theorem method'') based on the calculation of a universal distribution, which yields an objective and universal measure of algorithmic complexity for short strings that approximates Kolmogorov-Chaitin complexity.}

\shorttitle{Complexity for short strings}

\begin{document}
\maketitle                            
The production of randomness by humans requires high-level cognitive abilities such as sustained attention and inhibition, and is impaired by poor working memory. Unlike other frontal neuropsychological tests, random generation tasks possess specific features of interest: their demand on executive functions, especially inhibition processes, is high; and more importantly, training does not reduce this demand through automatization \cite{Towse07}. On the contrary, generating a random-like sequence requires continuous avoidance of any routine, thus preempting any automatized success. 

Random generation tasks have been widely used in the last few decades to assess working memory, especially (sustained) inhibitive abilities \cite{Miyake00}, in normal subjects as well as in patients suffering from a wide variety of pathologies. In normal subjects, random generation varies with personal characteristics or states, such as belief in the paranormal \cite{Brugger90} or cultural background \cite{Vandewiele86,Strenge09}. It affords insight into the cognitive effects of aging \cite{Heuer10}, hemispheric neglect \cite{Loetscher08}, schizophrenia \cite{Chan11}, aphasia \cite{Proios08}, and Down syndrome \cite{Rinehart06}. 

As a rule, random generation tasks involve generating a random-like sequence of digits \cite{Loetscher09}, nouns \cite{Heuer10}, words \cite{Taylor05} or heads-or-tails \cite{Hahn09}. Some authors have also offered a choice of more neutral items, such as dots, e.g., in the classical Mittenecker test \cite{Mittenecker58}. Formally, however, these cases all amount to producing sequences of bits, that is 0 or 1 digits, since any object can be coded in this way.
 In most studies, the sequence length lies between 5 and 50 items. Measuring the formal ``randomness" of a given short sequence (say of length 5 to 50) is thus a crucial challenge. Apart from any objective and formal definition of randomness, researchers regularly use a variety of indices, none of which is sufficient by itself because of the profound limitations they all exhibit. Recently for instance, Schulter, Mittenecker and Papousek \citeyear{Schulter10} provided software calculating the most widely used of such measures applied to the case of the Mittenecker Pointing Test, together with a comprehensive overview of the usual coefficients of randomness in behavioral and cognitive research. These tools provide a new way to describe how a given sequence differs from a truly random one. However, it is not fully satisfactory: multiple unsatisfactory measures do not result in a satisfactory description.


\section{The usual measures of randomness}
The most common coefficients used to assess the quality of a pseudo-random production may be classified into three large varieties according to their main focus.

\subsection{Departure from uniformity}
The simplest coefficients--even though they may rely upon sophisticated theories--are based on the mere distribution of outcomes, and are therefore independent of the order of the outcomes. In brief, they amount to the calculation of a distance between the observed distribution and the theoretical flat distribution, just as a chi-square would do.

 Information theory \cite{Shannon49} is often used as a basis for assessing randomness. Given a finite sequence $s$ of $N$ symbols repeatedly chosen from a set of $n$ elements, the average symbol information, also called entropy, is given by $ H(s)= -\sum_{i}p_{i} \log_{2}(p_i) $, where $p_i$ is the relative frequency of an item $i$ in the finite sequence. This entropy is maximal when the relative frequencies are all equal, when it amounts to $ H_{\max}=\log_2(n) $.

 Symbol redundancy (SR; see \citeNP{Schulter10}) is an example of a coefficient arising from information theory. It is defined to be $SR=1-H/H_{\max}$  where $H$ is the entropy. SR is no more than a measure of departure from uniformity. A sequence's SR does not depend on all aspects of the sequence, but only on the relative frequencies of each item comprising it. According to SR, a sequence comprised of 0s and 1s such as 000001 is weakly random as expected, but 010101, 000111 or 100101 are exactly equivalent to each other, since they all minimize SR to 0. This is the most obvious limitation of SR as a global measure of randomness, as well as of any measure relying on the mere distribution of symbols.

\subsection{Normality}
Beyond values depending on the distribution of individual symbols, one may consider pairs (dyads) or sequences of three (triads) adjacent outcomes. In a truly (infinite) random sequence, any dyad should appear with the same probability. Any distance between the distribution of dyads or triads (and so on) and uniformity may therefore be thought of as a measure of randomness. This is precisely what context redundancy $CR_1$ and coefficient of constraint $CC_1$ assess \cite{Schulter10}. 

One may also consider dyads of outcomes separated by 1, 2 or $k$ elements in the sequence, which is done, for instance, through $CC_k$ and $CR_k$ coefficients, a generalization of $CC_1$ and $CR_1$. Here we group methods of this kind under the rubric ``normality assessment" for a reason that will soon become clear.

 In mathematics, a sequence of digits $d_1,...,d_n,...,$ with $d_n$ lying within $[0,b]$ -- or equivalently, the real number written $0,d_1d_2d_3...$ in base $(b+1)$ -- is said to be normal in base $b$ if the asymptotic distribution of any dyad, triad, or of finite sequences of $k$ consecutive digits, is uniform. Any such series also manifests what may seem a stronger property: dyads of outcomes separated by a certain fixed number of other outcomes show, in the long run, a uniform distribution. Therefore, a normal sequence will be considered random by any normality assessment method. 

Eventually, there will exist sequences produced by simple rules that are normal. The Coperland-Erd\"os sequence, arising from the concatenation of prime numbers (235711131719...) is an example. The Champernowne sequence \cite{Champernowne33}, a concatenation of all integers from one on (123456789101112131415..., or in base 2 11011100101110111...), is an even more simple example. There even exist rules to generate absolutely normal sequences, i.e. numbers that are normal in any base $b$ \cite{Becher02}. Thus, sequences exist that would pass any normality test of randomness, despite being produced by a simple rule--which contravenes the notion that randomness is defined by an absence of regularity.

\subsection{Gaps}
Another variety of randomness coefficient is worked out using the rank distances between two identical items. For instance, in the sequence 12311, the distances between occurrences of the symbol 1 are 3 and 1.
 The 
frequency distribution of repetition distances (gaps) and the median of repetition gap distribution (MdG) are based on the study of the distance between two identical outcomes. They have proved useful in detecting the so-called cycling bias: people tend to repeat an item only after they have used every other available item once \cite{Ginsburg94}.

 Gap methods are as limited as normality assessment is: a normal sequence will pass these tests and be considered truly random, even if a naive rule  produces it.
 
 \subsection{Cognitive complexity}
 Apart from these attempts, a variety of measures to assess \emph{cognitive} complexity or randomness were put forward during the 60's and 70's (for an overview, see \citeNP{Simon72}).  These calibrations of cognitive complexity are built for the most part on the idea of algorithmic complexity, although in practice the link with Turing machines (Turing machine is at the root of Kolmogorov complexity theory) is evanescent. Usually, they consist of a series of rules or pattern descriptions that human subjects supposedly use and are aware of. The complexity of a string is defined as a function of the minimum number of rules one has to use to produce the string in question. These indices depend heavily on the rules and pattern descriptions chosen and lack mathematical rationale, but are psychologically sound. Their purpose is not to set a normative and formal measure of randomness through complexity, but to capture the nature of human complexity perception.
 
Very recently, a more sophisticated approach based on changes has been suggested \cite{Aksentijevic12}. Given a $n-$long string $s$, define the change function of $s$, $c(s)$, as the $(n-1)-$long string $f(s)$ whose $i-$th term is 0 if $s(i)=s(i+1)$, and 1 if not. From a string $s$ of length $n$, build a $n\times n$ matrix whose $j-$th line $L_{j}$ is $c(L_{j-1})$. Complexity is defined as a weighted sum of the matrix coefficients. Aksentijevic and Gibson (2012) claim that this \emph{change complexity} is an alternative to the unfortunately uncomputable algorithmic complexity.
 
However, there is no mathematical rationale for using change complexity as a normative measure of complexity or randomness. Rather than being a computable version of algorithmic complexity, it seems like a refinement of the usual normality assessment tools. It captures some local structural aspects of the string to be rated, and will consider every normal sequence as perfectly random, even though it is produced by a simple algorithm. Change complexity is also heavily reliant on psychological considerations and subjective choices. For these reasons, it should be classified with the previous attempts to capture \emph{perceived} randomness, but not as a formal measure of complexity. However useful it may be in capturing human perceived complexity or the production of pseudo-random strings, it is a bad candidate for rating human pseudo-random productions in a normative and objective fashion.

\subsection{Psychological justifications and limitations}
Notwithstanding their potential limitations, all the above-mentioned coefficients have proved useful in detecting some common biases in random generation. SR-like values capture outcome biases -- the overuse of certain symbols \cite{Nickerson02}. Normality assessment accurately spots alternation biases -- the tendency to avoid using the same symbol twice, e.g., HH or TT -- or the inverse \emph{repetition bias}. Context redundancy has also been linked with cognitive flexibility \cite{Stoffers01}, of which it constitutes an estimate. Gaps and related methods would diagnose the cycling bias, a tendency to repeat the same pattern or to exhaust every available symbol before a repetition \cite{Ginsburg94}. For instance, if the available symbols are 1, 2, 3, 4, a subject might choose 1, 3, 4, 2, 1, postponing the second appearance of ``1'' until after every symbol has occurred once. Repetition avoidance is known to affect outcomes as far as 6 ranks forward \cite{Chapanis95}, a bias that gap methods shed light on.

 It is unclear whether these measures happen to capture the basic biases in human random generation, or whether, unfortunately, researchers have focused on these biases simply because they have had tools at their disposal for diagnosing them. As we have seen in the previous sections, a normal sequence such as that suggested by Champernowne, which is highly non-random to the extent that it is generated by a simplistic rule, would meet all random criteria using symbol distribution, normality assessment, gap methods as well as change complexity.

At this point, we may list three senses in which the usual randomness estimates are unsatisfactory: First, they do not capture non-standard biases in randomness, such as the existence of a simple generation rule, when such a rule begins to produce sequences bearing some resemblance to a truly random one, e.g., a normal sequence. Only a few features of a random sequence are captured by these tailored measures. Second, they lack a theoretical basis. Despite being based on formal probabilistic properties, they are nevertheless not subsumed by a theory of randomness. In fact, they neither use nor provide any definition of a random sequence. Third, partly as an upshot of the first two points, several coefficients are needed to sketch an acceptable diagnosis of the quality of randomness, whereas a single measure would allow the comparison of sequences.

In what follows, we describe a method for overcoming the uncomputability of algorithmic complexity and thus arrive at a measure for short strings that possesses interesting features. First, it captures the complexity of a string in a single numerical value. This allows comparison between any two sequences, even if they are not of equal length. For instance, developmental studies of randomness (see below for a preliminary experiment) will benefit from having a global estimation of the quality of randomness. Second, because it is based on the Kolmogorov theory of randomness, it will detect any deviation from randomness or computable regularity, whatever the source of the discrepancy. And third, because it is linked with algorithmic probability (the probability that a string is produced by a random program), it serves as a bridge between complexity and the Bayesian approach. Knowing the  algorithmic complexity of a string $s$, one is able to compute the probability that it is produced by a random process, a result that would be meaningful within the new paradigm of subjective probability.

We must, however, underline the fact that the algorithmic complexity of short strings alone will not yield a description of any discrepancy detected between a theoretical and an experimental string. Investigating the manner in which human pseudo-random productions differ from truly random sequences (which is of course of great importance for psychology) will remain the domain of more specific measures focusing on particular characteristics of interest, such as the indices mentioned above.


\section{Algorithmic complexity}
The need for a universal approach that does not focus on specific arbitrary features, as well as for a theoretical framework defining finite randomness, has been expressed by psychologists \cite{Barbasz08} and addressed outside psychology by the mathematicians Andrei Kolmogorov and Gregory Chaitin.
 The theory of algorithmic complexity, also known as Kolmogorov-Chaitin complexity \cite{Li08}, provides a formal definition of a random sequence. In this section, we first provide an overview of this theory, identify its limits, and then suggest an approach for overcoming them, following recent developments in the field \cite{Delahaye07}.
 This approach is consistent with current theories of randomness perception in psychology. Falk and Konold have suggested that our subjective assessment of randomness relies on a perception of some sort of complexity \cite{Falk97}, thus linking perceived randomness and complexity. However, they do not provide any practical means of measuring randomness for short strings based on complexity. 

Our goal is to provide a new conceptual and practical tool enabling the assessment of randomness through algorithmic complexity for short strings within an objective and universal approach. It is thus not meant to capture human representations of randomness, but to verify that pseudo-random human behaviors can be rated on a numerical scale bridging randomness and complexity.

\subsection{Formal languages and automata theory}
When considering the complexity of an object, one may think of said object as simple if it can be described in a few words. One can, for example, describe a string of a million alternating zeros and ones 01010101...  as ``A million times 01'' and say that the string is simple given its short description. However, it is fair to point out that the description of something is highly dependent on the choice of language. The strings a language can compress depend on the language used, since any string (even a random-looking one) can be encoded using a one-word long description by mapping it onto any word of a suitable language. A language can always be tailor-made to describe any given object by using something that describes the object in a very simple way. Due to these difficulties it was not until the arrival of the theory of computation, and the precise definition of a computing machine by Alan Turing \cite{Turing36}, that the theory of algorithmic information found a formal framework on which it could build a definition of complexity and randomness.

 Today, the Turing machine model represents the basic framework underlying most concepts in computer science, including the definition of algorithmic randomness. A Turing machine (henceforth TM) is an abstract model of a digital computer formalizing and defining the idea of mechanical calculation.

 A TM consists of a list of rules capable of manipulating a contiguous list of cells (usually pictured as a tape with an unlimited quantity of cells in both directions), and an access pointer (an active cell) equipped with a reading head. The TM can be in any one of a finite set of states $Q$, numbered from 1 to $n$, with 1 the state at which the machine starts a computation. There is a distinct $n + 1$ state, called the halting state, at which the machine halts. Each tape cell can contain a 0 or a 1 (sometimes there is a special blank symbol filling the tape). Time is discrete and the instants of time (steps) are ordered from 0, 1, . . ., with 0 the time at which the machine starts its computation. At any given time, the head is positioned over a particular cell. The head can move right or left, reading the tape. At time 0 the head is situated over a particular cell on the tape called the initial cell, and the finite program starts in state 1. At time 0 the content of the tape is called the machine input.
 Turing's seminal contribution was the demonstration that there exist Turing machines (today named universal Turing machines, which we will denote simply by UTM) capable of simulating any other Turing machine \cite{Turing36}. One does not need specific computers to perform specific tasks; a single programmable computer could perform any conceivable task (we are now capable of running a word processor on the same digital computer on which we play chess games). He also proved that programs and data don't have any particular feature that distinguishes them from one another, given that a program can always be the input of another Turing machine, and data can always be embedded in a program. A full description of a Turing machine can be written in a 5-tuples notation as follows: $\{s_i, k_i, s_{i+1}, k_{i+1}, d\}$, where $s_i$ is the scanned symbol under the head, $k_i$ the state at time $t$, $s_{i+1}$ the symbol to write at time $t + 1$, $k_{i+1}$ the state to be in at time $t+1$ and $d$ the head movement either to the right or to the left. A TM can perform the following, which in combination define a single operation: (1) Write an element from $A = \{0, 1\}$. (2) Shift the head one cell to the left or right. (3) Change to state $k\in Q$. 

When the machine is running it executes one such operation at a time (one every step) until it reaches the halting state--if it ever does. At the end of a computation the TM will have produced an output described by the contiguous cells of the tape over which the head passed before halting. A TM may or may not halt, and if it does not halt it is considered to have produced no output. Turing also showed that there is no procedure to determine whether a Turing machine will halt or not \cite{Turing36}. This is set forth as the \emph{undecidability of the halting problem}, identified with the common term uncomputability.

\subsection{A definition of algorithmic complexity}
The basic idea at the root of algorithmic complexity is that a string is random (or complex) if it cannot be produced by a program much shorter in length than the string itself. The algorithmic complexity $C(s)$ of a bit string $s$ is formally defined as the length in bits of the shortest program $p$ that outputs it when running on a UTM $U$. Formally, $C(s) = \min_p\{|p| : U(p) = s\}$ , where $p$ is a program and $U(p)$ the corresponding output \cite{Kolmogorov65, Chaitin66}.

 One is compelled to use a universal Turing machine because one wants a machine that is capable of printing out any possible string $s$ for which one may want to calculate $C(s)$. Nevertheless, no computable procedure exists to calculate $C(s)$, due to Turing's undecidability of the halting problem, so $C(s)$ is usually approximated through the use of lossless compression algorithms, such as the Lempel-Ziv algorithm \cite{Lempel76}. The length of the compressed string $s$ is actually an upper bound of $C(s)$, hence a sufficient test of non-randomness. However, lossless compression algorithms do not help when strings are short -- shorter than, for example, the compression program length in bits. For instance, usual lossless compression algorithms give complexity above 40 for strings of length 1 or 2. Moreover, strange phenomena happen, such as the string ``11111'' being rated as far less complex than ``111'' or ``111111''. This and other arguments against the use of compression algorithms as estimates for complexity are detailed in a recent paper by Soler-Toscano, Zenil, Delahaye and Gauvrit (\citeyearNP{Soler12}, section 3.1 and Figure 1). In practice, a string of length $l\leq50$ cannot be compressed by lossless compression algorithms because the compressed file should include the decompression algorithm (otherwise one can compress anything to an arbitrary symbol with no further instructions), which usually needs more than 40 bits. Hence, Kolmogorov complexity by itself is of no use in the situations encountered by psychologists. This is probably the reason why, although algorithmic complexity has been successfully used in biology to assess the complexity of DNA sequences, it has, to the best of our knowledge, never been directly used in experimental psychology.

\subsection{The choice of Turing machine matters}
The definition of algorithmic complexity clearly seems to depend on the specific UTM $U$, and one may ask whether there exists a different UTM yielding different values for $C(s)$. The following theorem indicates that the definition of algorithmic complexity makes sense even if measured on different universal Turing machines (or if desired, using different programming languages):

\begin{theorem}[invariance]
If $U$ and $U^\prime$ are two UTMs and $C_U(s)$ and $C_U^\prime(s)$ the algorithmic complexity of a binary string $s$ using $U$ and $U^\prime$, there exists a constant $c$ that does not depend on $s$, such that for all binary strings $|C_U(s) - C_U^\prime(s)| < c$.
\end{theorem}

The proof of this theorem is quite straightforward--see e.g. \cite{Calude02}. The ability of universal machines to efficiently simulate each other implies that there is a program $p_1$ for the universal machine $U$ that allows $U$ to simulate $U^\prime$. One can think of $p_1$ as a translator in $U^\prime$ for $U$ (the length of $p_{1}$ is an upper bound for $c$). Let $p_2$ be the shortest program producing $s$ according to $U^\prime$. Then there is a program for $U$ made of $p_1$ and $p_2$ and generating $s$ as an output. 

For strings from a certain length on, this theorem indicates that one will asymptotically approach the same complexity value regardless of the choice of universal Turing machine. However, constants can make the calculation of $C(s)$ for \emph{short} strings profoundly dependent on the UTM used \cite{Delahaye08}. Both the problem of non-computability and the problems posed by short strings may account for the absence of applications in psychology and other social sciences.

\subsection{Algorithmic probability}
Solomonoff \citeyear{Solomonoff60} had the idea of describing the likelihood of a UTM generating a sequence with a randomly generated input program, having in mind a program to solve the problem of induction. Further formalizing this concept, Levin defined the \emph{algorithmic probability} of a string $s$ as the probability that a random program (the bits of which are produced with a fair coin flip) would produce $s$ when running on a UTM $U$ \cite{Zvonkin70}. Formally Levin's probability measure is defined as: $ m(s) = \sum_{\{p:U(p)=s\}} 1/2^{|p|}$. 

If $m(s)$ as a probability measure is not to be greater than 1, $U$ has, however, to be what is called a prefix-free Turing machine, that is, a machine that only accepts programs that are not the beginning of any other program (i.e. there exists an ``END'' sequence finishing any acceptable program). Levin's probability measure induces a distribution over programs producing $s$, assigning a greater chance that the shortest program is the one actually generating $s$.

\subsection{The Coding theorem method}
The \emph{Coding theorem} connects algorithmic probability (or frequency of production) to algorithmic complexity \cite{Calude02}:

\begin{theorem}[Coding theorem]
For any finite string $s$ and prefix-free universal Turing machine $U$,
$\mid - \log_2(m(s)) - C_U(s)\mid \leq M$, for a fixed number $M$ independent of $s$.
\end{theorem}

The Coding theorem offers an alternative for approximating $C(s)$, given that one can evaluate $C_U(s)$ by calculating $\log_2(m(s))$. This idea--the use of the Coding theorem as an alternative to compression algorithms as a means of estimating $C(s)$--has already been suggested by Delahaye and Zenil (2007). To this end, they sampled the space of all Turing machines up to a certain size (2 symbols and 4 states) from which a frequency distribution of the outputs generated by these machines was produced. We denote by $D(n)$ the probability distribution of all 2-symbol $n$-state Turing machines. In other words, $D(n)$ provides the production frequency of a string $s$, taking into account all 2-symbol $n$-state TMs. A string is counted in $D(n)$ if it is the output of a Turing machine, i.e. the content of the contiguous cells on the machine tape which the head has gone through before halting.

\subsection{Solving the halting problem for small machines}
$D(n)$, like $m(s)$ and $C(s)$, is a semi-computable function. One may use the results from a problem popular among computer scientists called the \emph{Busy Beaver problem} to calculate $D(n)$ as an approximation of $m(s)$. 

The Busy Beaver problem is the problem of finding a value $S(n)$ for the maximum runtime of a TM with $n$ states before halting, when starting from an empty input. Rad\`o \citeyear{Rado62} proves that finding $S(n)$ for any $n$ is impossible given the undecidability of the halting problem, but for $n < 5$ state and 2-symbol TMs, $S(n)$ is known. These values are arrived at by essentially simulating all different $n$-state machines and proving that the remaining non-halting machines will never halt (which is possible only because the machines are small and simple). Values of $S(n)$ for $n<5$ are known \cite{Shen65, Brady83}. These values of the Busy Beaver problem allowed Delahaye and Zenil \citeyear{Delahaye08,Delahaye12} to numerically approximate the complexity of strings using the Coding theorem\footnote{The tables produced by Zenil and Delahaye are available online at the following URL: www.algorithmicnature.org.}. In what follows, we will apply a similar procedure to the set of all 2-symbol 5-state Turing machines.


\section{$D(5)$ and $K_{5}$}
To date, $D(4)$ has been the most advanced distribution released. This paper represents a significant improvement both in accuracy and scope, providing new data from all 2-symbol 5-state Turing machines, denoted by $D(5)$. The following is the description of the procedure that we call the \emph{Coding Theorem Method}.

\subsection{Procedure} 
A Turing machine (TM) with $n$ states using the binary alphabet $\{0,1\}$ requires a set of $2n$ instructions, one for each possible $(state,symbol)$ pair. Each of these instructions provides a new state, a symbol to write on the tape, and a direction in which to move the machine head.

\begin{figure}[htbp!]
  \centering
  \subfigure[Transition table]{
    \label{examGrapRule}
  \includegraphics[width=.8\textwidth]{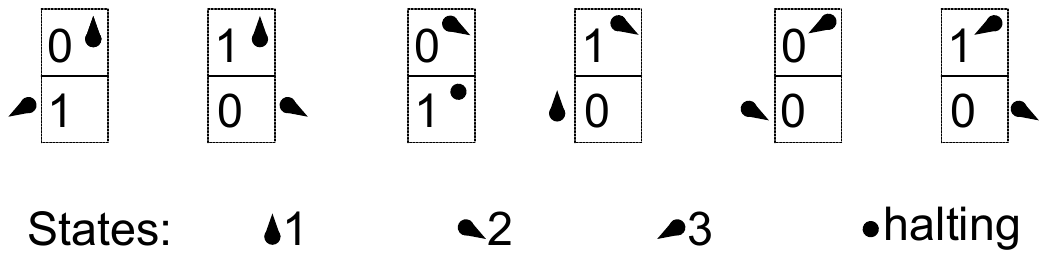}}
  \subfigure[Execution]{
    \label{fig:examGrapExec}\mbox{}\ \ 
  \includegraphics[width=.8\textwidth]{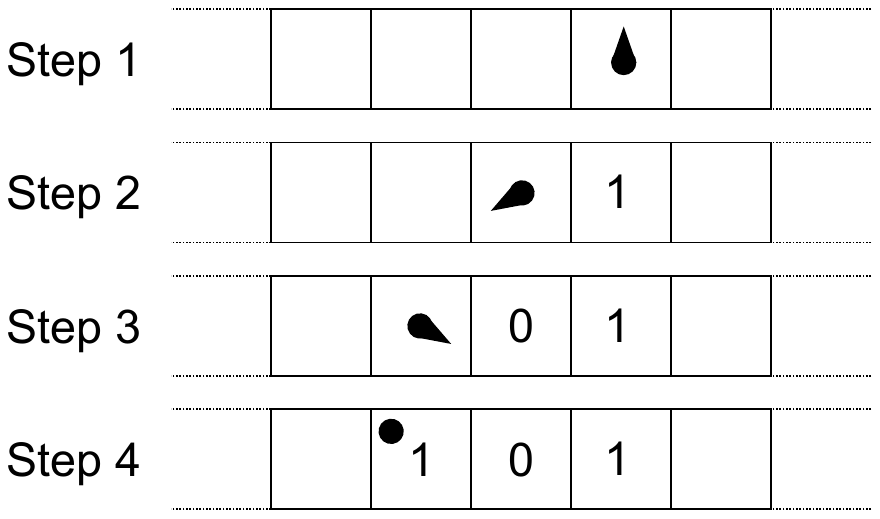}\ \ \mbox{}}
  \caption{An example of a machine with 3 states}
\end{figure}

Figure~\ref{examGrapRule} is a graphical representation of an
 example of a machine with three states. States are represented by black 
drops pointing in different directions (see legend). The 
halting state is represented by a black dot. The six two-cell 
pictures represent the instructions of the Turing machine. In each
 instruction, the top cell represents the state and the symbol read before 
applying the instruction. For example, the first instruction is 
applied when the Turing machine is in state 1 reading symbol `0'. The 
bottom part of each instruction represents the symbol written, the new
 state, and the direction (the new state is drawn to the right or left 
of the cell which is written on). So we can read the first instruction thus: when the 
machine is in state 1 reading the symbol `0', it replaces `0' 
with `1', moves to the left and changes to state 3. Note that when an 
instruction moves to the halting state (in 
Figure~\ref{examGrapRule} it only happens when the machine is in
state 2 reading `0') the head does not move, halting in the same
 cell.
Figure~\ref{fig:examGrapExec} displays the execution of this TM (on an unbounded tape). At the beginning, all the cells are blank. When the machine reads a blank, it identifies it as a 0. When the TM halts, the tape reads ``101'', which is therefore the string produced by this particular machine.

Each one of the $2n$ instructions in a Turing machine with $n$ states 
can be one of the $4n+2$ different combinations of
$(new\_state,new\_symbol,direction)$. When $new\_state$ is the halting 
state, there are only two possible instructions (writing `0' or
`1'). In other cases, there are $4n$ possibilities ($n$ possible values
 for $new\_state$, 2 for $new\_symbol$ and 2 for $direction$). That adds up to 
$4n+2$. Then, considering the $2n$ instructions of a Turing
 machine with $n$ states, we obtain the result that the number of different 
machines with $n$ states is $(4n+2)^{2n}$. 

Our experiment involved obtaining the output of all these machines 
for $n=5$, that is, $26\,559\,922\,791\,424$ different machines, machines starting on a blank tape filled with `0' as well as machines starting on a blank tape filled with
 `1' (in order to consider both symbols as possible blank 
symbols).

In fact, we did not run all the machines because we exploited some
 symmetries to considerably reduce the number of machines to run. For 
example, there is a left-right symmetry that allowed us to run only 
machines with the starting instruction moving in a fixed direction, 
for example, to the right. Then the output of some machines starting on 
the left is the reverse string of that produced by the left-right 
symmetric machines. There is also a `0'-`1' symmetry that made it unnecessary to run every machine twice, one for each possible blank 
symbol. We set `0' as the blank symbol and for every string produced 
we also considered its `0'-`1' symmetric string (interchanging `0' and
`1'). Also, we did not run many trivial machines that we knew in
advance would not halt (machines with the initial transition 
remaining in the initial state) or that would halt in just one step
(machines with the initial transition to the halting state). These
 reductions allowed us to run only $2(n-1)(4n+2)^{2n-1}$, which for $n=5$ 
meant taking up $4/11$ of the original space. 
At any rate, it took 18 days in the
 cluster at the CICA (Centro Inform\'atico Cient\'{\i}fico de
Andaluc\'{\i}a) supercomputing service, using a C++ simulator. 
Having obtained the output 
string of the machines that were run, we applied the proper completions to 
obtain the full set of strings representing the complete space of the 
different $(4n+2)^{2n}$ machines running both on `0'-filled and
 `1'-filled tapes.
 
This computation is very demanding, but only needs to be run once, which is now done.

\subsubsection{Setting the runtime}
The maximum number of steps that a Turing machine with 
5 states and 2 symbols can run before halting is still unknown. There are known 
machines running 47\,176\,870 steps, but we could not set that runtime 
for all the machines due to time limitations. So we decided to set a
 maximum runtime of 500 and keep only the output strings produced by 
machines requiring that number of steps at most to halt. 

\begin{figure}[htbp!]
  \centering
  \includegraphics[width=\textwidth]{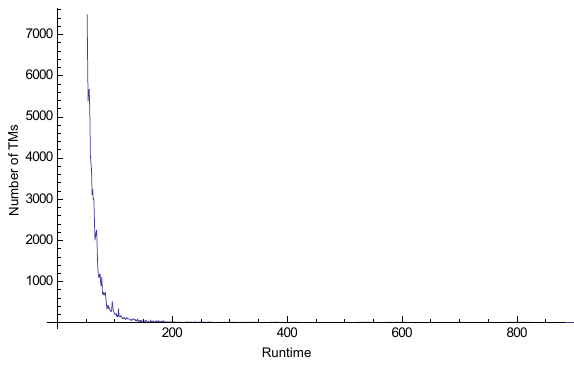}
  \caption{Runtime distribution in $D(5)$.}
  \label{fig:runtimeD5}
\end{figure}

As an experiment to try and ascertain how many machines we were leaving out, we ran $1.23 \times 10^{10}$ random Turing machines for up to 5000 steps. Figure~\ref{fig:runtimeD5} gives the number of TMs which halt after exactly $n$ steps. The rapid decrease denotes the fact that very few machines actually halt after more than 200 steps. Among these, only 50 machines halted after 500 steps and before 5000. Thus according to this experiment, our election of a  runtime of 500 steps leaves out only an insignificant portion of the  machines.

Not all machines were run for 500 steps, as we set some filters to 
detect non-halting machines before consuming the maximum runtime. For
example, cycles of period two are easy to detect. They are produced 
when in steps $s$ and $s+2$ the tape is identical and the machine is
 in the same state and the same position. When this is the case, the 
cycle will be repeated infinitely.

\subsubsection{Calculating $D(5)$}

The output of our experiment is the list of strings produced by 
machines with 5 states and 2 symbols, together with the number of
 occurrences of each string. We proceeded to construct the function $D(5)(s)$, 
which returns the probability that a halting machine with 5 states and 2 
symbols produces string $s$ when running on a blank tape (without 
specifying the blank symbol). Formally,
\[D(5)(s) = \frac{\text{number of occurrences of }s}{\text{number of
 halting executions}}\]
Note that the number of halting executions is equal to the sum of 
occurrences of all the strings produced, so the sum of $D(5)$ for all strings is 1.

 \subsection{Features of $K_{5}$}
 Function $D(5)(s)$ is a natural approximation of $m(s)$, from which an approximation of Kolmogorov-Chaitin complexity may be derived by use of the Coding theorem. Formally, we define the $D(5)-$based $K_{5}$ algorithmic complexity for short strings by the formula
 $$ K_{5}(s)=-\log_{2}(D(5)(s)).$$
 
Every string of length $l\leq11$ was a 5-state 2-symbol TM output, as were 4094 of the 4096 strings of length 12, and some strings up to length 49, making for a total of 99608 strings.

\subsubsection{Examples of top and bottom strings}
Table \ref{tab:simple} displays the 12 most frequent, and therefore least complex strings of length 12 according to $D(5)$, as well as the top 12 least frequent (hence most complex) strings. It confirms our expectations: strings with apparent regularity  (``000000000000'' or ``010101010101'') are simple, and the more complex ones exhibit no such visibly regular pattern.

\begin{table}[tbp]
\caption{Top 12 simple strings according to $D(5)$ (top of the table) and 12 more complex strings (bottom).}
\label{tab:simple}
\begin{tabular}{ccc}
\hline
111111111111 & 000000000000 & 101010101010\\
010101010101 & 111111111110 & 100000000000\\
011111111111 & 000000000001 & 110101010101\\
101010101011 & 010101010100 & 001010101010\\
\hline
\hline
011000111001 & 000100110111 & 111000110100\\
110100111000 & 001011000111 & 000111001011\\
110100011100 & 110001110100 & 001110001011\\
001011100011 & 110000111100 & 001111000011\\
\hline
\end{tabular}
\end{table}

\subsubsection{Detection of global regularities}
Some binary sequences may seem simple from a global viewpoint because they show symmetry (1011 1101) or repetition (1011 1011). 
Let us consider the string $s=1011$ as an example. We have $D(5)(s)=3.267414\times 10^{-3}$. The repetitions $ss=10111011$ have a much lower probability $D(5)(ss)=4.645999\times 10^{-7}$. This is not surprising considering the fact that $ss$ is much longer than $s$, but we may then wish to consider other strings based on $s$. In what follows, we will consider three methods (repetition, symmetrization, 0-complementation). The repetition of $s$ is $ss=10111011$, the``symmetrized" $s\bar{s}=1011 1101$, and the 0-complementation $10110000$. These three strings of the same length have different probabilities ($4.645999.10^{-7}$, $5.335785\times 10^{-7}$ and $3.649934\times 10^{-7}$ respectively). 

\begin{figure}[h!]
\includegraphics[width=13cm]{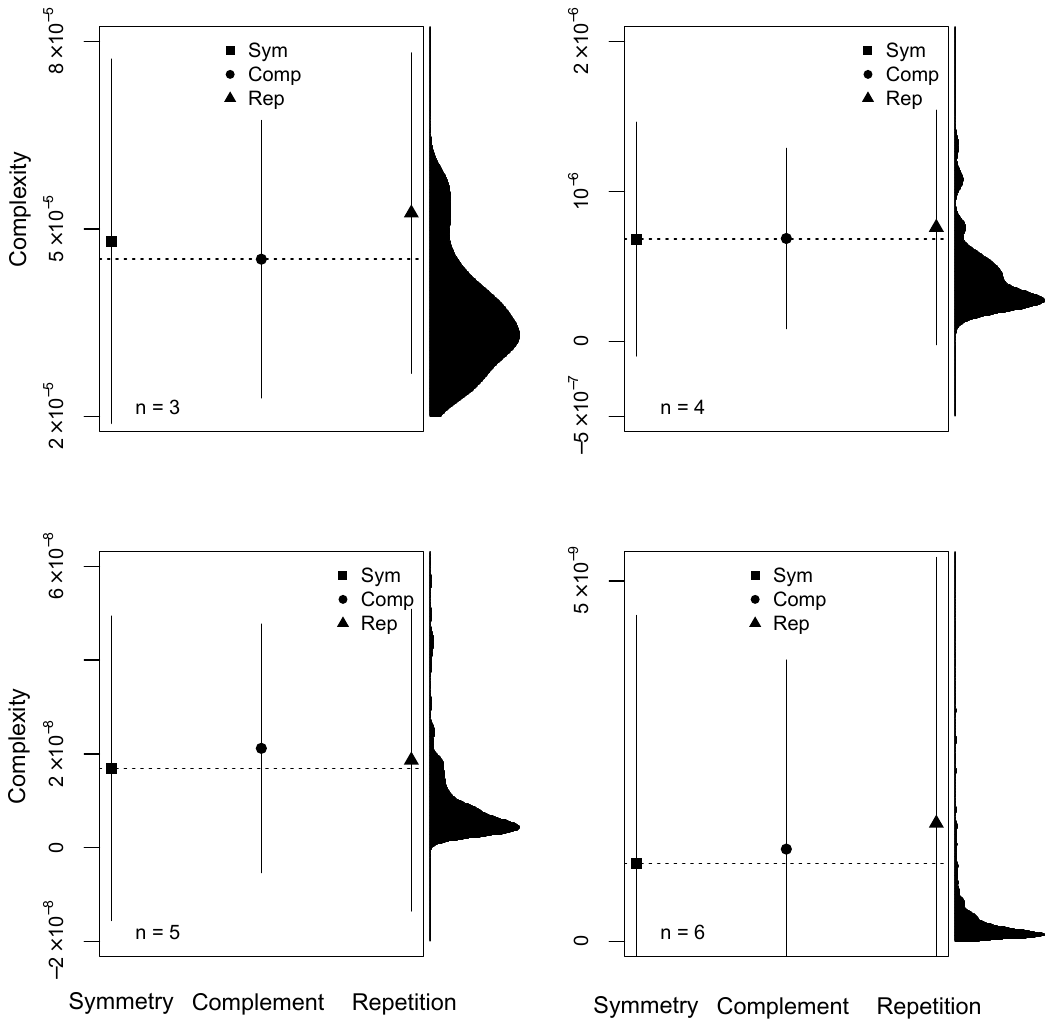}
\caption{Mean $\pm$ standard deviation of $D(5)$ of $2n$-long strings given by processes of symmetrization (Sym), 0-complementation (Comp) and repetition (Rep) of all $n$-long strings. The dotted horizontal line shows the minimum mean among Sym, Comp and Rep. The density of $D(5)$ for all $2n$-long strings is given in the right-margin.}
\label{SymCompRep} 
\end{figure}

Let us now consider all strings of length 3 to 6, and their symmetrizations, 0-complementations and repetitions. Figure \ref{SymCompRep} provides a visual representation of the results. In each case, even the minimum mean between the means of symmetrized, complemented and repeated patterns (dotted horizontal line) lies in the upper tail of the $D(5)$ distribution for $2n$-length strings, and this is even more obvious with longer strings. Symmetry, complementation and repetition are, on average, recognized by $D(5)$.

\subsubsection{Alternations}
Human pseudo-random productions have been described as exhibiting too much alternation: when trying to behave randomly, humans have a proclivity to produce `1' after `0' or `0' after `1'. The mean frequency of alternation in a random binary string is .5, but slightly superior frequencies (typically around .6) have been reported in human pseudo-random generation (e.g. \citeNP{Tubau09}).

 It is now widely believed that when human subjects try to behave randomly, they actually try to maximize the complexity of their responses. However, due to cognitive limitations, we probably are unable to produce binary sequences of maximal complexity, because this would require a too complicated algorithm. This theoretical intuition recently received experimental support when researchers found that children were more attracted by mildly complex patterns \cite{Kidd12}.

\begin{figure}[h!]
\includegraphics[width=13cm]{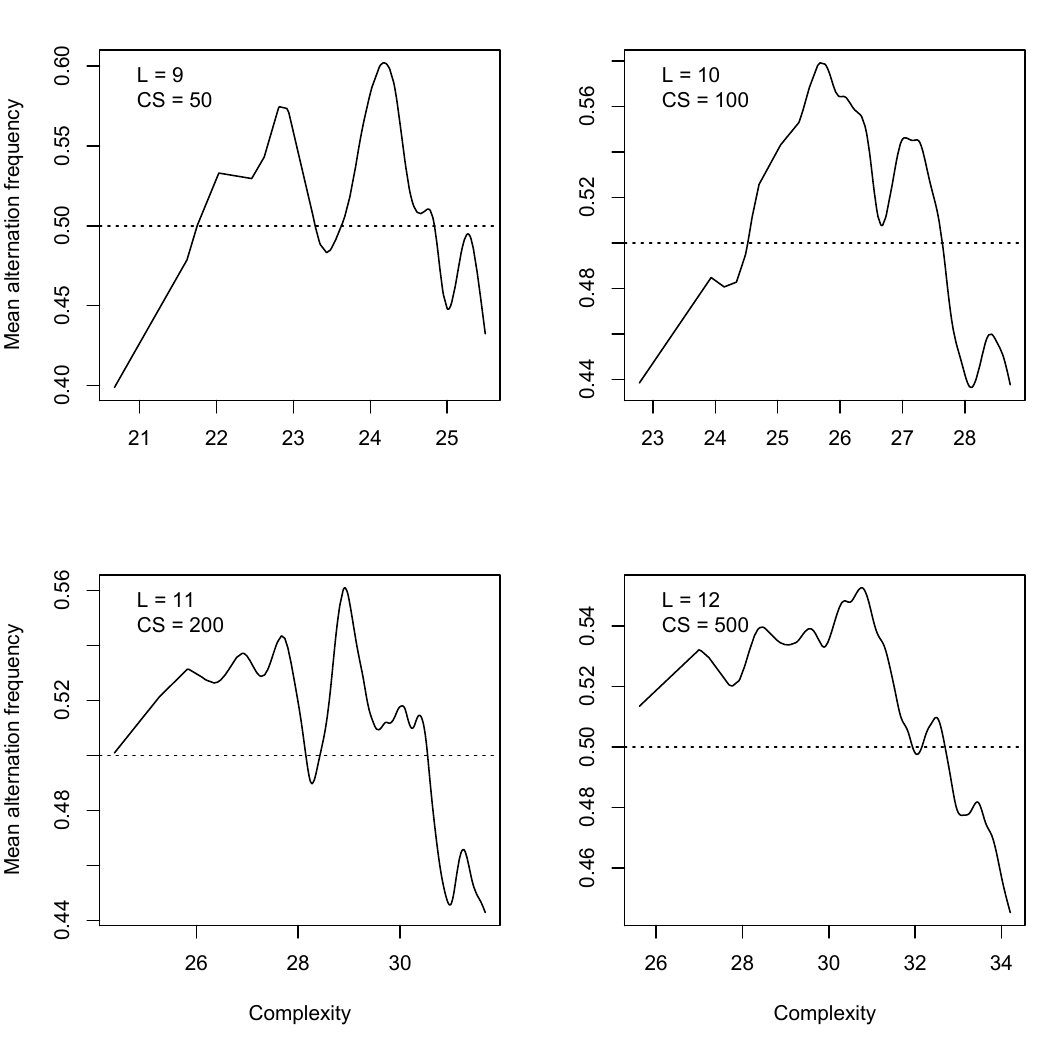}
\caption{For each length $l=9$ to $12$, we select clusters of $CS$ strings and compute the mean frequency of alternations within these strings. The diagram displays the smooth-spline curve of the resulting function. The number $CS$ of strings is chosen according to the total number of strings of length $l$, to ensure readability. The clusters are built thus: strings are sorted by increasing complexity. For a given complexity $k$, let $j$ be the rank of the first string with complexity above $k$. Then the corresponding cluster comprises all strings with rank $j$ to $j+CS-1$.}
\label{alternations} 
\end{figure}

Figure \ref{alternations} shows that binary strings of medium or mildly high complexity tend to exhibit an excess of alternations, whereas both simple and very complex strings tend to have an excess of repetitions. Complexity is a means to a deeper understanding of alternation bias. Formally, we suggest the following conjecture: When human subjects try to behave randomly and produce a binary string, they try to generate a complex sequence. Because their efforts are only partially rewarded, they produce strings of mild or just-above medium complexity, which usually have too much alternation.

 \subsubsection{Bayesian use of $D(5)$}
 Given a string $s$, we may now compute the probability of it being truly random (event $R$) against the hypothesis that it has been created by a TM (event $M$). 
Let $l$ be a length (for instance, $l=12$). In this paragraph, we will denote the conditional probability of $D(5)$ when the length is $l$ by $P_{l}$.
 Set the prior probability that the underlying process is random to $P(R )=1/2$. We then have $P(s|R)=2^{-l}$ and
$$P(R|s)=\frac{P(s|R)P(R)}{P(s)},$$
with $$P(s)=P(s|M)P(M)+P(s|R)P(R)=\frac{P_{l}(s)}{2}+{2^{-(l+1)}}.$$ 
From this we derive $$P(R|s)=\frac{1}{1+2^{l}P_{l}(s)}.$$

\begin{figure}[h!]
\includegraphics[width=13cm]{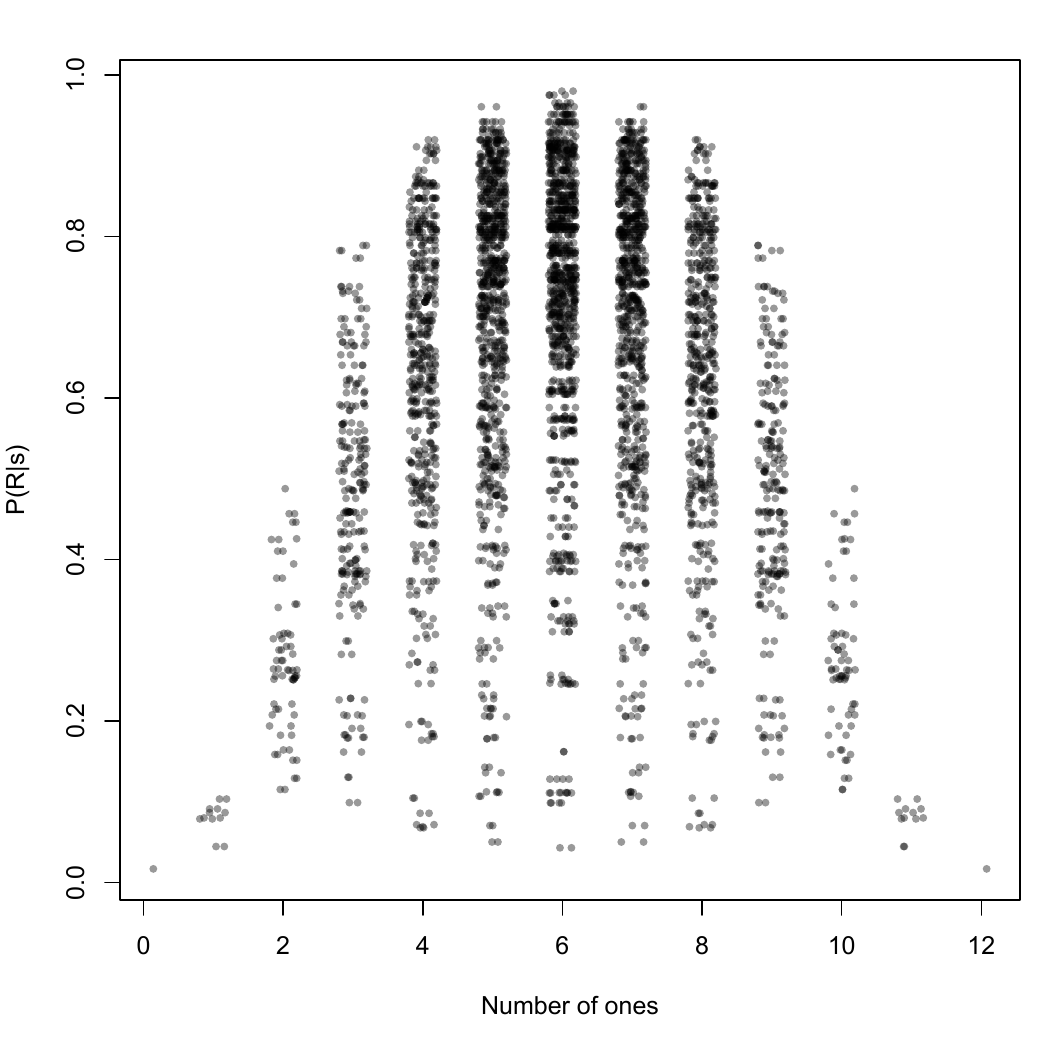}
\caption{Scatterplot of the probability of being random against the number of 1s in all $12-$long strings. A jitter method has been applied on the number of 1s to increase readability.}
\label{CompPR} 
\end{figure}

Table \ref{tab:tab2} gives a few examples of random binary strings of length 12 together with their probability of being random, and the number of 1s they include. As we can see from this table and as figure \ref{CompPR} visually confirms, the more complex strings (which also have a great probability of being random) are likely to be balanced, with approximately six 1s. This again may help us understand the so-called ``belief in the law of small numbers'' \cite{Tversky71} according to which subjects wrongly tend to generate equal numbers of each alternative while trying to produce random binary strings. Once again, this could be a result of an attempt to generate complex responses, that is, strings that are more likely to be randomÉ

\begin{table}[tbp]
\caption{A few examples of $12-$long strings $s$ with the associated probability of being random $P(R|s)$ and the number of 1s (6 meaning that $s$ is perfectly balanced).}
\label{tab:tab2}
\begin{tabular}{ccc}
\hline
$s$ & $P(R|s)$ & Ones\\
\hline
110100011100 & 0.975 & 6\\
101100100011 & 0.951 & 6\\
110001111001 & 0.942 & 7\\
101100111100 & 0.933 & 7\\
100000000011 & 0.519 & 3\\
110101001011 & 0.515 & 7\\
101110110111 & 0.180 & 9\\
101111101111 & 0.151 & 10\\
010101010101 & 0.097 & 6\\
010000000000 & 0.079 & 1\\
000000000000 & 0.017 & 0\\
\hline
\end{tabular}
\end{table}

\subsection{Material and tools}
An Online Algorithmic Complexity Calculator (OACC) implementing the technique presented herein to approximate the Kolmogorov complexity of short strings and meant for the use of the research community is accessible at \url{http://www.complexitycalculator.com}. The online calculator provides a means for approximating the complexity of binary short strings for which no other method has existed hitherto by taking advantage of the formal connections among these measures and putting together several concepts and results from theoretical computer science.

 The calculator is a long-term project to develop an online tool implementing the semi-computable measures of complexity described in this paper, and is expected to be expanded in the future. It currently implements numerical approximations of Kolmogorov complexity and algorithmic probability for short binary strings following the numerical methods described herein, strings where lossless compression algorithms fail when used to approximate their Kolmogorov complexity. Hence it provides a complementary and alternative method to compression algorithms. 

The OACC is intended to provide a comprehensive framework of universal mathematical measures of randomness, structure and simplicity for researchers and professionals. It can be used to provide objective measures of complexity in a very wide range of disciplines, from bioinformatics to psychometrics, from linguistics to finance. More measures, more data and better approximations will be gradually incorporated in the future, covering a wider range of objects, such as longer binary strings, non-binary strings and $n$-dimensional arrays (such as images).

The raw data set with the full calculation of $D(5)$ can also be downloaded from \url{http://complexitycalculator.com/D5.csv} in Comma Separated Values format. For R-users, a script is also available. It defines a function returning the complexity of any binary string up to length 11, and many strings of length 12 to 49. It also computes the algorithmic probability $D(5)$ and the probability of being random in a Bayesian approach $P(R|s)$. This R-script may be used to compute complexity in data frames, thus allowing a statistical exploration of complexity.

\subsection{A comparison with entropy}
Let us now briefly compare algorithmic complexity as defined above and the most widespread alternative measure of randomness: entropy. We will focus on binary strings of length 12. There are $2^{12}=4096$ different binary strings of length 12.

Among the 4096 binary strings of length 12, entropy distinguishes 6 groups. It is not defined for strings ``000000000000'' and ``111111111111'', and for any other string it only depends on the number of the least frequent digit, which can appear 1, 2, 3, 4, 5 or 6 times.

Algorithmic complexity is far more fine-grained, distinguishing 371 different levels of complexity.

The class of maximum entropy ($H=-\log_{2}(2)=1$) includes 461 different strings, of varying algorithmic complexity. Within this group, we find  strings of very low complexity such as ``010101010101'' ($K_{5}=26.99$) and of high complexity, such as ``100101110001'' ($K_{5}=36.06$).

Some strings of lower entropy sometimes have greater $K_{5}$ than others with higher entropy. For instance, the string ``110011011111'' has an entropy of 0.811, 4th among 6 possible level of entropy, but a complexity of $K_{5}=33.36$. Thus, ``110011011111'' is less random than ``010101010101'' according to its entropy, but more random according to its $K_{5}$. All the strings of very low entropy ($H < 0.5$) also have low complexity, but both simple and complex strings can
have medium or high entropy. 

On a more theoretical level, the Kolmogorov definition of a random string is equivalent to another mathematical construct, \emph{effective statistical tests}: a string is random if and only if it passes every effective statistical test. An important mathematical advantage of the theory of algorithmic complexity is that it constitutes the only tool that would detect any computable regularity \cite{martin66}. The measures mentioned above (entropy, contextual redundancy, change complexity, gaps, etc.) may be thought of as particular computable statistical tests of randomness. They each aim at detecting one variety of non-randomness. On the other hand, the Kolmogorov approach is equivalent to running every computable statistical test at the same time. In a way, it can be thought of as a universal statistical test of randomness.


\section{A case study}
In this short section, we present and briefly discuss experimental data gathered via an Algorithmic-Complexity-for-Short-Strings approach, for illustrative purposes. This study is part of a larger experiment still in progress that investigates the evolution of subjective probability with age in children and teenagers.

\subsection{Subjects and procedure}
68 children (35 male, 33 female) aged 7 to 11 participated in this experiment (mean age $\pm$ SD: $8,7\pm1,1$). All the children were pupils in 5 different classes from 2 public schools in France, grades 2 to 4. The youngest subjects (2d graders) were the first to be tested, followed 4 months later by the 3rd-graders, and 4 months after that by the 4th graders. 

Each child was received individually in a room in the school, during class time. A token was presented to him or her. It was green on one side and red on the other. The experimenter explained what was meant by a ``toss'' and the token was then hidden so it would not distract the subject. 

The instruction given to the child was to imagine that s/he tossed the token 8 times, and to say out loud which side, green or red, appeared each time. Each child thus produces one binary string that is 8 units long.

\subsection{Global complexity}
The complexity of all possible strings of length 8 runs from 18.53 to 22.68, with a median of 21.60. The more frequent strings generated by the children are 00101101, 01001101, and 11010010, appearing 4 times each. These strings have medial complexities (21.58, 21.37 and 21.58 respectively). 

On the whole however, children do better than a true random process in terms of complexity. The mean complexity of a random string of length 8 is 21.46, whereas children's productions have a mean complexity of 21.74 (SD: 0.52). This is significantly more than 21.46 ($t(67)=4.35, p<5\times 10^{-5}$). Of course, choosing a random Turing machine instead of a random string would lead to an over-representation of simple sequences, hence children's productions are also ``better'' than random Turing machine outputs.

\begin{figure}[h!]
\includegraphics[width=13cm]{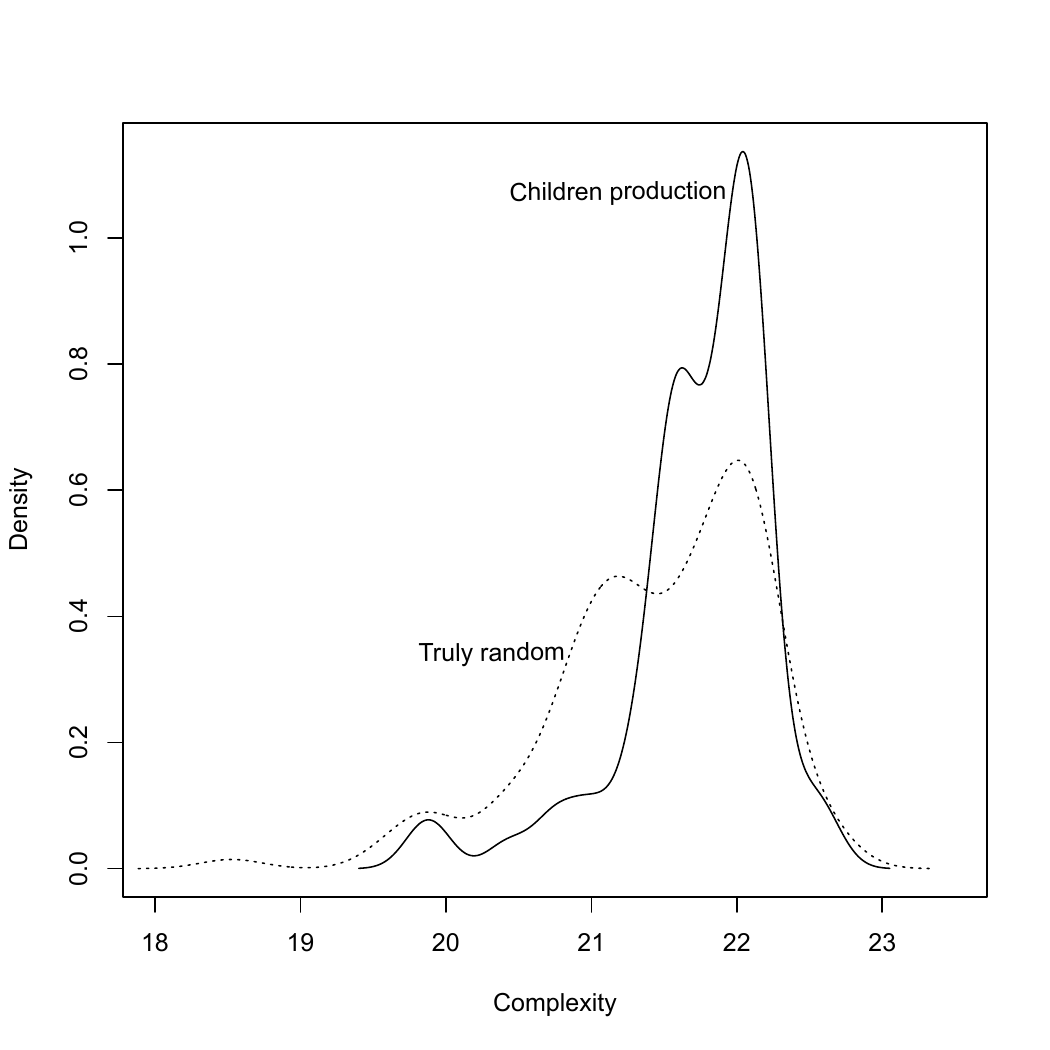}
\caption{Density of truly random strings' complexity (dotted) and of children's productions, as approximated by Gaussian kernel.}
\label{Density} 
\end{figure}

Figure \ref{Density} shows the density of complexities of truly random strings and of those produced by children. It strongly suggests that the main difference between truly random strings (i.e. every string shares the same probability of being picked up) and human production is that humans contrive to avoid very simple strings. They do not generate more high-complexity sequences than expected from chance, but they do produce a lot of mildly complex strings.

\subsection{Effect of education}
The period from 7 to 11 years is known to be one of relative stability. Piaget and Inhelder \citeyear{Piaget51} claimed that the notion of probability couldn't be grasped before the formal stage, at around age 11. In an experimental study of randomness perception in 7 to 16 year-olds, Green \citeyear{Green86} finds no evolution between 7 and 11. 

\begin{figure}[h!]
\includegraphics[width=13cm]{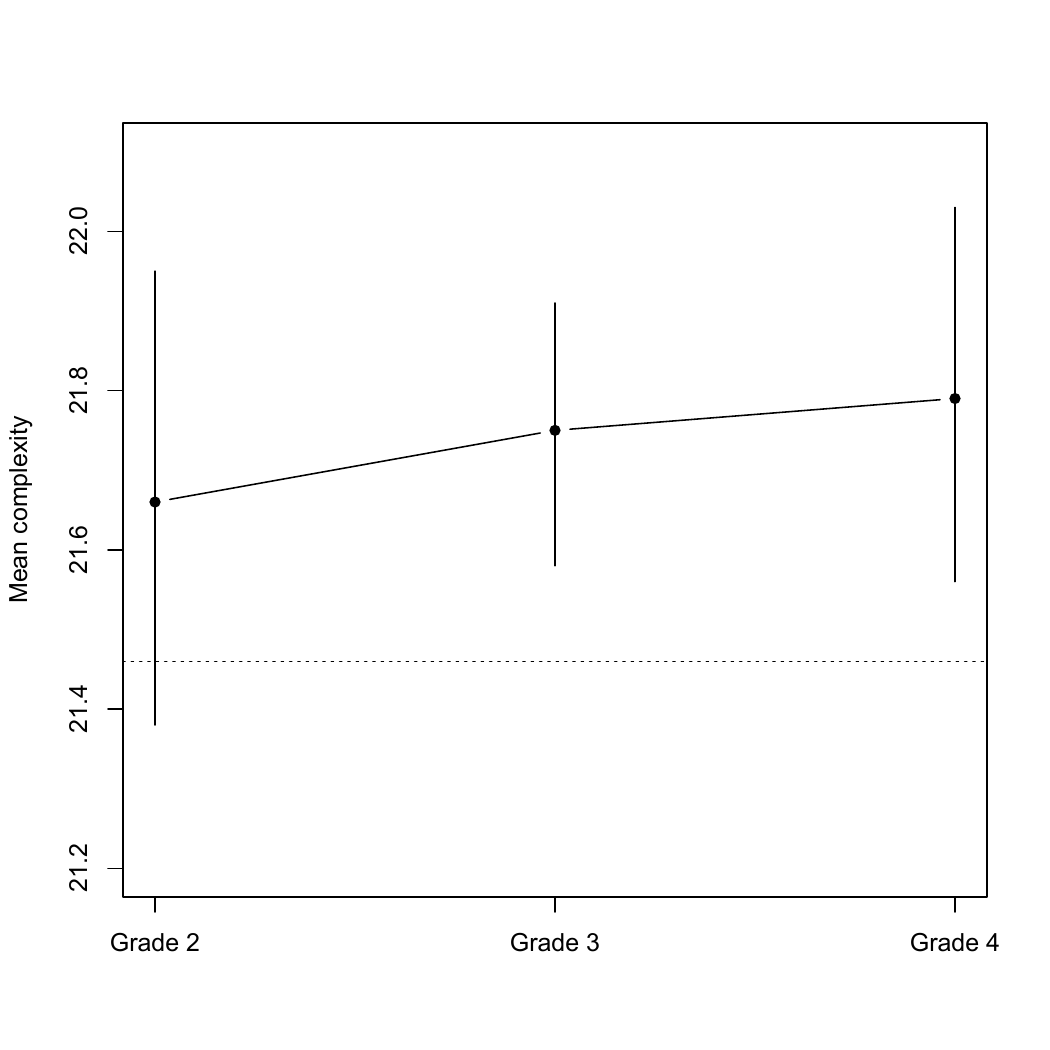}
\caption{Mean complexity $K_{5}$ and 95\% confidence intervals by grade. The dotted line shows the mean complexity of truly random binary strings of length 8.}
\label{ComplexityByGrade} 
\end{figure}

One aim of the present experiment is to ascertain whether there is an evolution in subjective randomness at this particularly stable stage. Concerning complexity, we expected either no differences from grade 2 through grade 4, or an increase in complexity. Figure \ref{ComplexityByGrade} displays the means and confidence intervals by grades. A one-way anova was applied to the data. Although there is no significant difference $(F(2,65)=0.68, p=0.41)$, a small effect \cite{Cohen92} ($\eta^{2}=0.11$) appears in our sample and shows a slight progression of complexity with grade.

The main finding of this experiment is that even young children aged 7 to 11, from grade 2 to 4, can do better than picking a string from a uniform distribution to build up strings that ``look random''. This behavior is not an irrational bias: any string is certainly as likely to appear by chance, but not all of them show the same amount of evidence for a random underlying process. The possible slight evolution of complexity between grades 2 to 4 confirms that this period of childhood is relatively stable.


\section{Conclusion}
For decades, researchers have relied upon algorithmic Kolmogorov-Chaitin complexity as a means to rate complexity and randomness in an objective way. Algorithmic complexity may seem to be a purely theoretical apparatus, useful only in abstract operations on infinite sequences. It has turned out to be useful in practice when applied to DNA sequences, or when long strings (of millions of digits, for instance) are involved. In such cases, compression methods seem to overcome the impossibility of an exact calculation of algorithmic complexity. Compression methods unfortunately don't apply to short strings, which we encounter in the behavioral sciences. Since amending the theory to make it suitable for short strings was believed to be impossible, researchers have followed two routes:

\begin{enumerate}
\item Some renounced the ideal of a normative unique measure of complexity, and used a variety of indices focusing on special features of the strings they had to measure.
\item Others switched from universal Turing machines to tailor-made cognitive ``machines'' designed according to what they thought was important for humans. In doing so, they generated good estimates of \emph{perceived} complexity, but lost the universality of algorithmic complexity.
\end{enumerate}

The Coding theorem method described above allows us to use algorithmic complexity even for very short strings. As we have shown, this may shed new light on some biases such as the alternation bias or the belief in the law of small numbers: they may be simple consequences of a tendency to be mildly complex. Because $K_{5}$ measures an objective complexity, we cannot expect human productions to be as complex as the more complex possible strings, though a mild and above average complexity may be expected.

 As Peter Brugger puts it\footnote{Personal communication, 6th November 2012}, until now ``a complexity measure for short strings [was] badly needed''.

\bibliography{NGauvritbib}
\end{document}